# Decoherence-free measurements in steering tasks

Artur Szczepański [1]

Institute of Fundamental Technological Research, Warszawa, Poland

Quantum entanglement has been used as an experimental tool to control the state of a quantum subsystem by measurements in its twin subsystem by Kwiat and Chiao 21 years ago, already, with entangled subsystems linked by an auxiliary classical coincidence channel. We propose to extend the experiment one significant step further by eliminating the coincidence channel. The current availability of bright biphoton sources makes such experiments feasible.

The scheme of the Kwiat and Chiao [1], (KC), control device is shown on Fig. 1.:

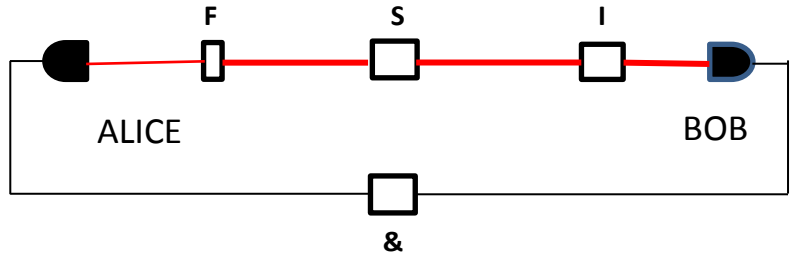

Fig. 1. Abridged scheme of the KC experiment. Idler and signal energy-entangled photons generated by **S** are directed in the ALICE's and BOB's spatial modes, respectively. ALICE uses the filter **F** to modify the spectral content of idler photons reaching her detector. BOB observes the interference pattern generated by the signal photons in the interferometer **I** in coincidence with ALICE's detections. **&** is the coincidence logic in the classical channel linking ALICE and BOB.

The setting of the KC experiment resulted in observed fringes when a narrowband filter **F** has been used, and in a flat interference pattern with a broadband **F**.

According to my note of 1992 [2], (AS), the outcome of KC could be used to evaluate experimentally "the propagation velocity of the collapse effect, $\kappa$" understood in AS as the quotient,

$$\kappa = \frac{space\ distance\ between\ \mathbf{F}\ and\ BOB's\ detector}{time\ of\ fight\ between\ \mathbf{F}\ and\ BOB's\ detector} \quad . \quad (1).$$

The denominator, that is the "time of flight, $\tau$, of the collapse effect" has been inferred from the photon's propagation velocity in the relevant optical path, and the threshold distance $|F - S|_T$ beyond which the collapse effect generated by **F** is no longer seen by BOB. The occurrence of the threshold distance $|\mathbf{F} - \mathbf{S}|_\mathbf{T}$ has been assumed in AS. Note however, that this assumption can be tested in the considered scheme just by varying $|\mathbf{F} - \mathbf{S}|$.

_________________________

[1]). Most of the work has been done at the Institute, but I am retired now and my actual address is: artallszczep@gmail.com

If we define

$$\tau = (|\mathbf{DB} - \mathbf{S}| - |\mathbf{F} - \mathbf{S}|_T) c^{-1} , \qquad (2)$$

with $|\mathbf{DB} - \mathbf{S}|$ as the length of the optical path from the source to BOB's detector, and $c$ the photon velocity along the considered path, then $\kappa$ would be given in measurable terms.

The "speed of quantum information", a quantity corresponding to $\kappa$, has been measured by the Gisin group (GG) in experiments with a quantum channel different from that in KC, but with a classical coincidence channel linking ALICE and BOB as in KC, [3-5]. More specifically the "speed's" lower bound has been evaluated as exceeding $c$ by several orders of magnitude.

At this point a meta-theoretical principle of status-consistency could be evoked:

The parts (components) of an object, the status of which is physical, are physical too. The parts of a theoretical formal object (e.g. constructs of a physical theory, mathematical objects, etc) have a formal theoretical status. Hence, for example, an object with parts of different status would be inconsistent. Accordingly, we should not expect to observe experimentally objects containing non-physical components (e.g. formal), and we should not expect to be able to construct consistent theoretical objects containing real physical components.

Now, consider a task involving a classical action and a quantum action as components of the complete action. If we are compelled to ascribe the physical status to the complete action (its result), or to its classical part (classical partial result), the status-consistency would require the same physical status for the quantum part of the action (quantum partial result). Specifically, this requirement could apply to the quantum information understood here as the information transferred by means of the quantum channel in the KC experiment.

Relevant experimental arguments are provided by GG, if the "speed of quantum information" is interpreted as measured in a real, physical procedure (see, however, the discussion [6,7]).

Thus, if we do not reject the status-consistency principle, we could admit the following conjecture:

The status of quantum information in experiments is physical. (PSC)

Taking seriously PSC would imply involvement in the extremely delicate matter of quantum nonlocality consequences. On the other hand, both recent theoretical [8,9] (see, also the comment [10]), and experimental (see, e.g. [11-16]) arguments strongly encourage gathering experimental evidence aimed directly at the problem.

Here I propose some conceptually simple experimental schemes aimed at testing PSC. The former can be regarded as a direct consequence of AS: The idea is to remake the experiments proposed in [2] **without** the coincidence circuit, which should be technically feasible due to the current availability of bright biphoton sources (see. e.g. [17], and the sources used in experiments [11–16]). To be more specific consider the schemes below.

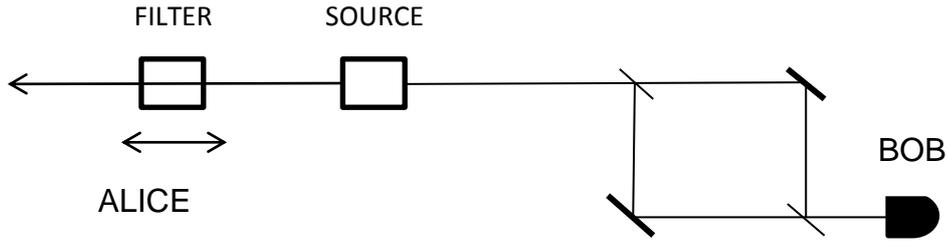

Fig. 2. Scheme of the steering experiment with energy-entangled biphotons. The A-photon is emitted by the SOURCE into the Alice's spatial mode, the twin B-photon goes into Bob's spatial mode. Alice controls the length of the optical path between the SOURCE and the FILTER, which cuts off a part of the A-photon's frequency spectrum (projective action). Bob's detector sees the interference pattern (decohering-free measurement).

Suppose Alice to set the optical path length SOURCE – FILTER (|**S - F**|) exceeding the optical path length SOURCE – Bob's detector (|**S - BD**|):

$$|\mathbf{S - F}| > |\mathbf{S - BD}|. \quad (3)$$

The entangled system evolves unitarily before the BD action, and Bob sees an interference pattern as resulting from the emitted B-photon's wave packet frequency spectrum and the specific experimental conditions.

Now, let

$$|\mathbf{S - F}| < |\mathbf{S - BD}|. \quad (4)$$

The decohering action of F on the A-photon, which precedes the BD action, consists in either the transmission of the A-photon, or the absorption of the A-photon. Both the cases result in modification of the B-photon spectrum, and consequently Bob would observe a modified interference pattern as reported in [1]. Note that without the coincidence circuit the observed modified pattern would consist of the addition (superposition) of the pattern generated by photons transmitted by F and the of the pattern generated by photons absorbed by F.
This would be the manifestation of the predicted **steering** (or spooky action at a distance) effect.

Bob's measurement does not affect the A-photon's frequency spectrum, which is the physical result of the F and the BD action formal non-commutation, and which implies the one-way of the steering effect in the considered scheme.

Consider now the polarization-entangled photons version:

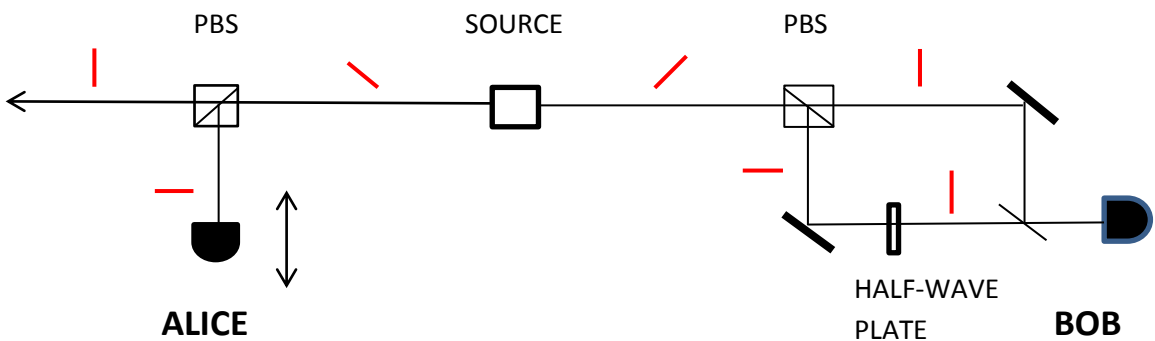

Fig. 3. Scheme of the steering experiment with polarization-entangled photons. The linearly polarized A-photon and B-photon is split by Alice's and Bob's polarizing beam

splitter (PBS) into vertical and horizontal components, respectively. The half-wave plate in the interferometer restores the polarization co-planarity of the split B-photon's parts, so that Bob can see fringes in the interference pattern. Alice controls the optical path length between the source and her detector. The transmitted part of the A-photon propagates freely. Red bars exemplify the polarization planes of the beams.

Suppose first, as in the previous scheme, that

$$|S - AD| > |S - BD|. \qquad (5)$$

The system evolves unitarily before **BD**'s action, and Bob sees fringes in the interference pattern.

Now, if

$$|S - AD| < |S - BD|, \qquad (6)$$

the (projective) action of **AD** would absorb or not the A-photon, i.e. one of the split A-photon's part would vanish. Consequently, steering allows only one of the B-photon split parts to survive at **BD**, so the interference pattern seen by **BD** is flat, which is the **steering effect** manifestation. Note that **AD** acts as a black screen, since its measurement result is irrelevant to the meaning of Bob's outcome.

Both the proposed schemes would give reliable results provided the source is bright enough to grant a sufficient signal/noise ratio. A modified scheme with one-wing heralded photons would provide an enhanced signal/noise ratio (see, Fig. 4 below). The modification does not affect the expected in the previous schemes main results i.e. the **classical channel-free steering effect**, and the capability to evaluate the lower bound of its "**propagation speed**".

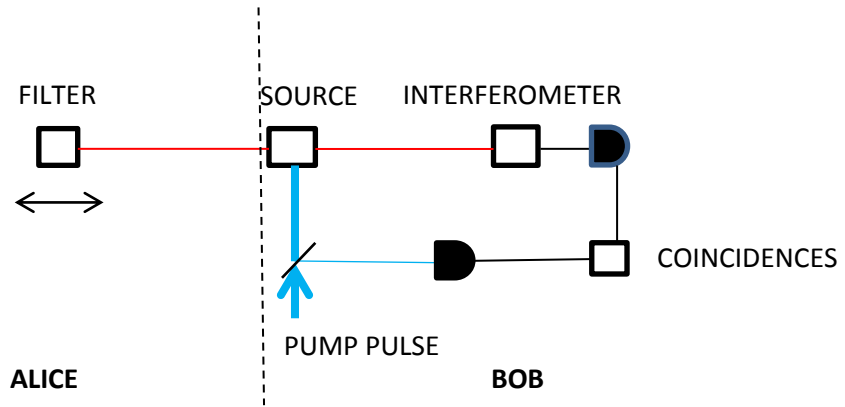

Fig. 4. Scheme of experiments with heralded B-photons. Alice operates the length of the optical path between the SOURCE and her projective-acting component (here exemplified by the FILTER). Bob operates the PUMP PULSES, the SOURCE, and observes the interference pattern of B-photons heralded by the pump pulses. Interference measurements do not decohere the entangled biphoton system, so the A-photons beam does not contain information about Bob's photon selective measurement. The scheme can be used in the polarization-entanglement experiment of Fig. 2.

In conclusion, the extension of experiments proposed in [2] should open the way for new tests of classical-channel-free steering tasks taking advantage of the noncommutation of projective and decoherence-free measurements in entangled subsystems. The expected results should provide strong arguments in the foundation question concerning the physical status of quantum information transfer and related quantum objects. On the other hand, any rational conclusions regarding the general consequences of the expected experimental outcomes would be involved in unavoidable metatheory considerations. But the latter are beyond the scope of this note, which has been limited to indicating new schemes of steering experiments.

**Acknowledgments**

Numerous helpful discussions with Robert Gałązka (both private and at the Institute of Physics, PAS, Warsaw) are gratefully acknowledged. I would like to thank Jakub Rembieliński and the participants of the seminary at the Chair of Theoretical Physics, University of Łódź, and the participants of the "Open Systems" seminary at the Department of Mathematical Methods in Physics, Warsaw University, for discussions.